# Ambient Effects on Photogating in MoS$_2$ Photodetectors

*Peize Han\*, Eli R. Adler, Yijing Liu, Luke St. Marie, Abdel El Fatimy$^†$, Scott Melis, Edward Van Keuren and Paola Barbara\**

Department of Physics, Georgetown University, Washington, DC 20057, USA.

\*e-mail: ph523@georgetown.edu; paola.barbara@georgetown.edu

ABSTRACT: Atomically thin transition metal dichalcogenides (TMDs) are ideal candidates for ultrathin optoelectronics that is flexible and semitransparent. Photodetectors based on TMDs show remarkable performance, with responsivity and detectivity higher than $10^3$ AW$^{-1}$ and $10^{12}$ Jones, respectively, but they are plagued by response times as slow as several tens of seconds. Although it is well established that gas adsorbates such as water and oxygen create charge traps and significantly increase both the responsivity and the response time, the underlying mechanism is still unclear. Here we study the influence of adsorbates on MoS$_2$ photodetectors under ambient conditions, vacuum and illumination at different wavelengths. We show that, for wavelengths sufficiently short to excite electron-hole pairs in the MoS$_2$, light illumination causes desorption of water and oxygen molecules. The change in the molecular gating provided by the physisorbed molecules is the dominant contribution to the device photoresponse in ambient conditions.

KEYWORDS: Two-dimensional materials, hysteresis, photodetectors, photoresponse

MANUSCRIPT TEXT:

Monolayer transition metal dichalcogenides (TMDs), such as $MoS_2$, $WS_2$, $MoSe_2$ and $WSe_2$, have attracted great interest for optoelectronic applications because, even with their drastically reduced thickness, they interact with the incident light strongly due to their direct bandgap and large density of states[1]. For example, phototransistors based on exfoliated flakes of monolayer $MoS_2$ show excellent performance, with responsivity as high as $10^4$ $AW^{-1}$, leading to shot-noise-limited detectivity larger than $10^{13}$ Jones [2]. Similar results were recently demonstrated with phototransistors based on monolayer $MoS_2$ grown by chemical vapor deposition (CVD) [3]. However, applications are hindered by the slow response time, typically as long as tens of seconds [4, 5]. Several previous studies showed that the environment strongly affects the properties of TMD transistors [6-10], including their photoresponsivity and speed [2, 11]. This is expected since all the $MoS_2$ molecules are on the surface of the material and any other molecules adsorbed from the environment can create charge traps [1]. For both CVD-grown and exfoliated $MoS_2$ transistors, the hysteresis in the measurements of source-drain current as a function of gate voltage is a typical signature of water and oxygen molecules adsorbed on the $MoS_2$ surface or between the $MoS_2$ and the substrate [5, 6, 12]. Encapsulation of $MoS_2$ devices with $Si_3N_4$ [6] and $HfO_2$ [2] partially removes the water and oxygen molecules with a substantial reduction of the hysteresis and increase of the mobility. Similar results are obtained when the devices are measured under vacuum [10]. Furthermore, the hysteresis will increase as humidity increases [6]. All these experiments on the effect of ambient conditions on the hysteresis can be explained with the adsorption (desorption) of oxygen and water molecules on the $MoS_2$ surface that occurs when a positive (negative) gate bias is applied [7], as we will discuss later. However, the effect of ambient conditions on the photoresponse of $MoS_2$ devices is still unclear.

The responsivity of a photodetector is the ratio between the photocurrent and the power of incident light, where the photocurrent is the difference between the current measured when the device is illuminated and current measured when the device is not illuminated (dark current). There are two mechanisms contributing to the photocurrent: photoconduction and photogating [13]. Photoconduction is due to the free charge carriers generated by absorption of light with photon energy larger than the bandgap of $MoS_2$. If all the photons are absorbed, and each incident photon generates one electron and one hole, the maximum of photoresponsivity to light with 600-nm wavelength is about 1 $AW^{-1}$, which is orders of magnitude smaller than the responsivities typically measured with $MoS_2$ photodetectors. This indicates that the dominant mechanism of photoresponse in $MoS_2$ phototransistors is photogating, where light absorption causes a change in the density of trapped charges. Since trapped charges change the effective gate voltage, when photogating occurs there is a large increase of current, due to a shift of the threshold voltage $V_{TH}$, the gate voltage separating the high-current (On) and low-current (Off) regimes in the $MoS_2$ transistor. In this work, we study the underlying mechanism of the photogating, its relation to the presence of water and oxygen molecules and to the hysteresis of $MoS_2$ phototransistors.

The samples discussed here were fabricated following the fabrication process reported in our previous work [3]. In short, a CVD-grown $MoS_2$ film was patterned as a 200 μm by 60 μm slab on the growth substrate by photolithography (using a PMMA/SU8 bilayer photoresist [14]) and deep reactive ion etching (DRIE). The $MoS_2$ slab was then transferred to a heavily p-doped Si chip capped with 300 nm $SiO_2$. Although it has been shown that wet transfer typically creates sulfur or molybdenum vacancies that can serve as adsorption sites[15], we use a wet transfer that does not degrade the optical properties of the $MoS_2$, as described in our previous work[3]. The source and drain electrodes were also patterned by photolithography. We fabricated devices with graphene electrodes and Cr(2 nm)/Au(250 nm) electrodes. These two types of devices exhibited similar electrical characteristics and photoresponse [3]. Figure1 (a) shows a typical $MoS_2$ device with gold electrodes and Figure 1(b) shows the photoresponse from one of those devices, with no gate voltage applied.

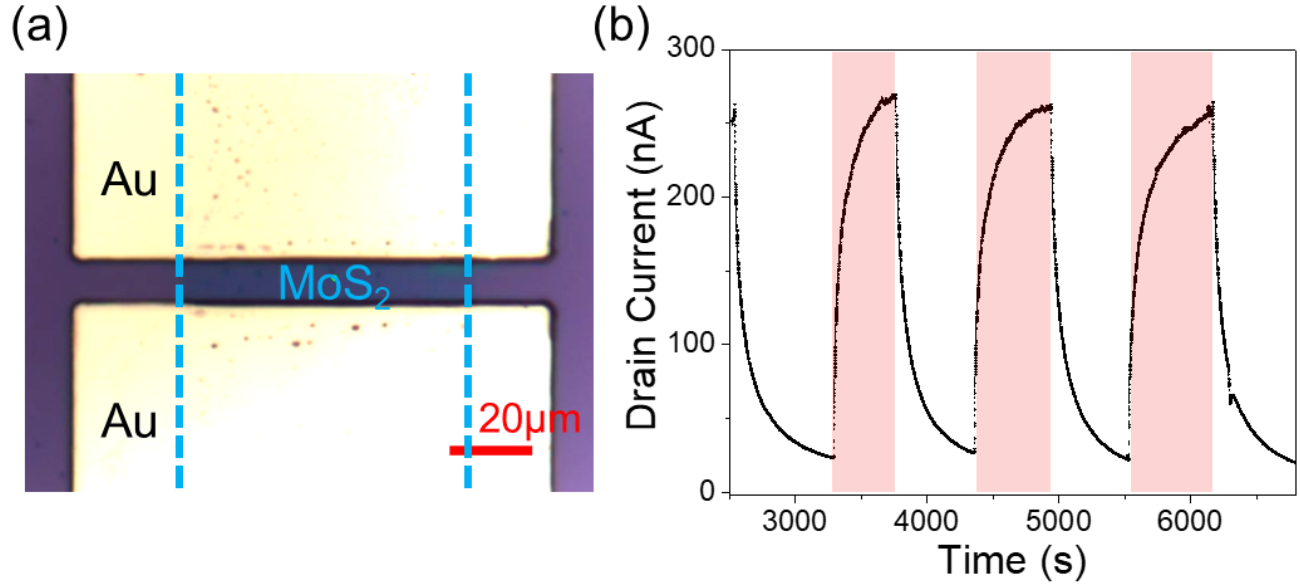

**Figure 1.** (a) Optical image of a $MoS_2$ device with gold electrodes. The area between the blue dashed lines is the $MoS_2$ slab (b) Time-resolved photoresponse with source-drain voltage $V_{SD}$ = 4V, and gate voltage $V_G$ = 0V. The red regions show the time intervals when the 633-nm laser is on. The power density of the laser is 50 µW $cm^{-2}$.

The photoresponsivity is about 800 $AW^{-1}$, under irradiation of an expanded laser spot, with wavelength of 633 nm and power density of 50 µW $cm^{-2}$. Similar to other $MoS_2$ devices from exfoliated or CVD-grown samples measured under ambient conditions, the response is very slow [3, 4, 16]. For example, the decay regime of the time-dependent photoresponse curve can be fit by the sum of two exponentials with different time constants, y = $A_1$*exp (-x / $\tau_1$) + $A_2$*exp (-x / $\tau_2$) + $y_0$ [17]. Typical values for $\tau_1$ and $\tau_2$ are tens and hundreds of seconds, respectively. The slow photoresponse has been attributed to the fact that the samples are exposed to ambient conditions[2]. Kufer and Kostantatos showed that encapsulation of the devices with $HfO_2$ substantially reduces the response time by three orders of magnitude[2]. At the same time, the responsivity is also reduced by about three orders of magnitude [2]. These results suggest that the same mechanism responsible for the slow response time is also responsible for the high responsivity. The encapsulation of the devices with $HfO_2$ strongly reduced the devices' hysteresis, confirming that it had partially removed water and oxygen molecules from the $MoS_2$ surface, as discussed above. Our devices are not encapsulated. They are measured in ambient condition and, as expected, they exhibit a substantial hysteresis of the source-drain current, $I_{SD}$, measured as a function of the gate voltage $V_G$, as shown in Figure 2(b). The mechanism leading to the hysteresis is explained by Cho et al. [7] and sketched in Figures 2(a) and 2(c).

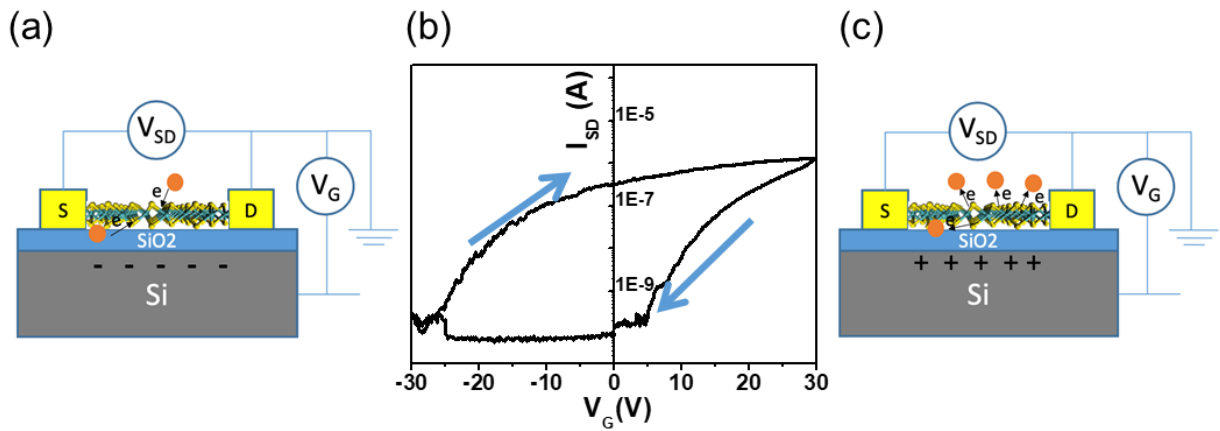

**Figure 2.** Charge transfer between the water molecules (orange dots) and the $MoS_2$ at (a) $V_G$ = -30V and (c) $V_G$ = 30V in the $I_{SD}$-$V_G$ plot (b).

When a positive gate voltage is applied, the electron concentration in the $MoS_2$ increases (Figure 2(c)). Electrons are easily captured by water and oxygen molecules and the concentration of adsorbed molecules on the $MoS_2$ surface increases [7]. Since the molecules are removing electrons from the $MoS_2$, they are effectively hole-doping the channel and shifting the threshold voltage towards positive gate voltages (to the right, in the plot in Figure 2 (b)). At negative gate voltages, the $MoS_2$ channel is filled with holes as shown in Figure 2(a) and the electrons are transferred back from the water and oxygen molecules to the $MoS_2$ to recombine with the holes ($O_2^- + h \rightarrow O_2$ and $H_2O^- + h \rightarrow H_2O$), while the water and oxygen molecules desorb [7, 11] or stay confined between the $MoS_2$ and the substrate as empty electron traps. Therefore, at negative gate voltages, the concentration of adsorbed molecules and electron-filled traps (along with their hole-doping effect) is lowest and the threshold shifts towards negative gate voltages (to the left in Figure 2 (b)). A threshold at negative gate voltages is indeed expected for "pristine" (without adsorbed molecules) $MoS_2$, because CVD-grown $MoS_2$ is electron doped. Density functional theory (DFT) calculations confirm this picture[18]. DFT predicts that physisorption of $O_2$ and $H_2O$ molecules on $MoS_2$ occurs with binding energies of 79 meV and 110 meV, respectively, and that approximately 0.04 electrons per $O_2$ and 0.01 electrons per $H_2O$ are transferred to the molecules[18].

To distinguish between the effect of the water confined at the interface between $MoS_2$ and substrate and the water adsorbed and desorbed at the top surface of the $MoS_2$, we compared the devices fabricated on $Al_2O_3$ and $SiO_2$ surfaces. Since $Al_2O_3$ is hydrophobic and $SiO_2$ is hydrophilic (after the piranha cleaning required for graphene transfer), we expect a smaller concentration of water molecules confined between the $MoS_2$ and the $Al_2O_3$. As both devices were measured in ambient conditions, the density of adsorbates on the top surface should be similar to each other, therefore the difference of the hysteresis between the two devices can be assumed to be due to the different concentration of water molecules confined between the $MoS_2$ and the substrate.

For the $MoS_2$ on the $Al_2O_3$ surface, the device was fabricated after depositing 30-nm of $Al_2O_3$ by atomic layer deposition on the p-doped Si chip capped with 300-nm $SiO_2$ (the pre-patterned CVD-grown $MoS_2$ slab was transferred on $Al_2O_3$ and the Ti(2nm)/Au(250nm) electrodes were patterned by e-beam lithography).

As shown in Figure 3, the hysteresis of the device on $SiO_2$ is much larger than the hysteresis of the device on $Al_2O_3$, confirming that the water trapped between the $MoS_2$ and the substrate contributes substantially to the hysteresis.

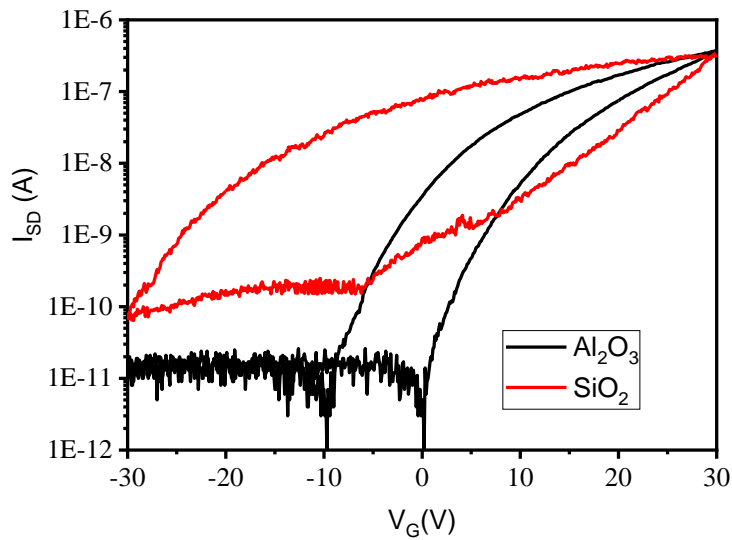

**Figure 3.** The $I_{SD}$-$V_G$ plot of the $MoS_2$ devices on $SiO_2$ and $Al_2O_3$.

The device fabricated on $Al_2O_3$ also shows lower source-drain current below the threshold gate voltage and a larger On-Off ratio. This is consistent with the work discussed by Ahn *et al.*[10], showing that reduced adsorbates can lower the Schottky barrier for electrons, causing an increase of current above the threshold gate voltage. This shift in the band alignment will in turn cause an increased Schottky barrier for holes, with smaller source-drain current at negative values of gate voltage and a larger On-Off ratio[10].

While hydrophobic substrates are an effective way to reduce water molecules between the bottom surface of the $MoS_2$ and the substrate, vacuum pumping is an effective way to reduce the concentration of molecules adsorbed on the top surface. Several works showed that the hysteresis is substantially reduced under vacuum conditions [6, 7, 10]. There are also other effective ways to remove water and oxygen molecules from the top surface of $MoS_2$. Miller et al. studied the Raman spectrum of $MoS_2$ in ambient conditions as a function of light intensity [19]. They argued that the laser illumination causes the desorption of water molecules, thereby changing the effective doping of the $MoS_2$ and causing a red shift of the Raman $A_{1g}$ mode [19]. They also argued that the molecular gating from the adsorbates is reversible and can be precisely tuned with the light intensity[19].

Following these results, we measured the effect of vacuum and illumination on the hysteresis of our $MoS_2$ devices. Figure 4 shows the current vs. gate voltage for one of our devices fabricated on $SiO_2$ with graphene electrodes and measured under different conditions.

The black curve is the measurement in ambient conditions. The blue curve shows the measurement in vacuum, at $10^{-5}$ Torr, with the hysteresis substantially reduced, as expected from the removal of some oxygen and water molecules. The red curve is the measurement during the laser illumination (633nm, 1mW cm$^{-2}$) while pumping, to hinder re-adsorption of the molecules removed by the laser illumination from the MoS$_2$ surface. The green curve shows the measurement done after switching the laser off, with the device still in vacuum. The hysteresis is smaller and the threshold has shifted to more negative gate voltage values compared to the (blue) curve measured in vacuum before illumination. During illumination, since the sample is kept under vacuum while pumping, the water molecules desorbed by the light are pumped away, therefore, once the illumination is turned off, the density of molecules re-adsorbed on the surface is smaller than it was prior to illumination. This is consistent with the smaller hysteresis and the threshold shift toward negative gate voltage values that we measure for the green curve and shows that the combination of vacuum and illumination is more effective in removing adsorbed molecules than vacuum alone. Similar to previous work [10], we noticed that the mobility increases as the number of adsorbates is reduced (the mobility in ambient conditions is 0.1 cm$^2$V$^{-1}$s$^{-1}$, the mobility in vacuum is 0.3 cm$^2$V$^{-1}$s$^{-1}$ and the mobility in vacuum after laser illumination is 0.6 cm$^2$V$^{-1}$s$^{-1}$). The increased mobility under vacuum can be explained by the reduced concentration of scattering centers from the adsorbates and the modulation of the Schottky barriers at the interface between the MoS$_2$ and the contacts, yielding a higher current above threshold[10].

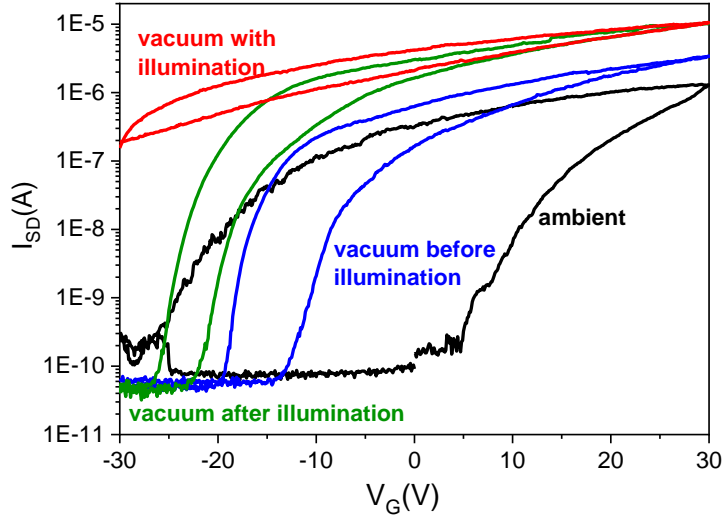

**Figure 4.** The comparison of $I_{SD}$-$V_G$ curves of a $MoS_2$ device on $SiO_2$ measured in ambient (black), vacuum (blue), vacuum with illumination (red) and vacuum after illumination (green).

There can be different ways in which the laser-assisted desorption can occur. One possibility is laser cleaning, with a cleaning rate that depends on the light intensity [18, 19]. Another possibility is that the laser generates electron-hole pairs in the $MoS_2$ and facilitates the desorption of oxygen and water molecules from the $MoS_2$ surface through the reaction $O_2^- + h \rightarrow O_2$ and $H_2O^- + h \rightarrow H_2O$, similar to the effect of a negative gate voltage [7, 11] discussed earlier. This process only occurs with photon energy sufficiently high to create electron-hole pairs.

To clarify the mechanism of laser-assisted desorption, we measured the $I_{SD}$-$V_G$ curves of the gold-contacted $MoS_2$ device on the $Al_2O_3$ substrate with illumination at different wavelengths, as shown in Figure 5(b). (Before the measurement, the device was annealed in 100 sccm Ar and 10 sccm $H_2$ at 200 °C for 2 hours to improve the contact [20]). A Xenon light with a monochromator was used as a wavelength-tunable light source (see Methods section).

Figure 5 (b) shows that a clear threshold shift toward negative gate voltage only occurs for wavelengths shorter than 650 nm, corresponding to photon energy higher than 1.9 eV. Figure 5(a) shows the photoluminescence spectrum of $MoS_2$ on $Al_2O_3$. The A and B peaks are due to excitonic transitions and their energy separation arises from the splitting of the valence band, due to spin-orbit coupling[21]. The A and B peaks are at about 1.85 eV and 2.0 eV, respectively, consistent with other work[22]. Our measurements show that the substantial threshold shift (photogating) occurs for photons energies higher than the A exciton peak, indicating that the water and oxygen desorption processes are caused by the photogenerated holes from the A excitons. We note that this photogating effect occurs for photon energies smaller than the bandgap, therefore photogain mechanisms based on the generation of free charge carriers and on the difference between the majority carrier transit time and the transit time of the minority carrier [16, 23] can be ruled out.

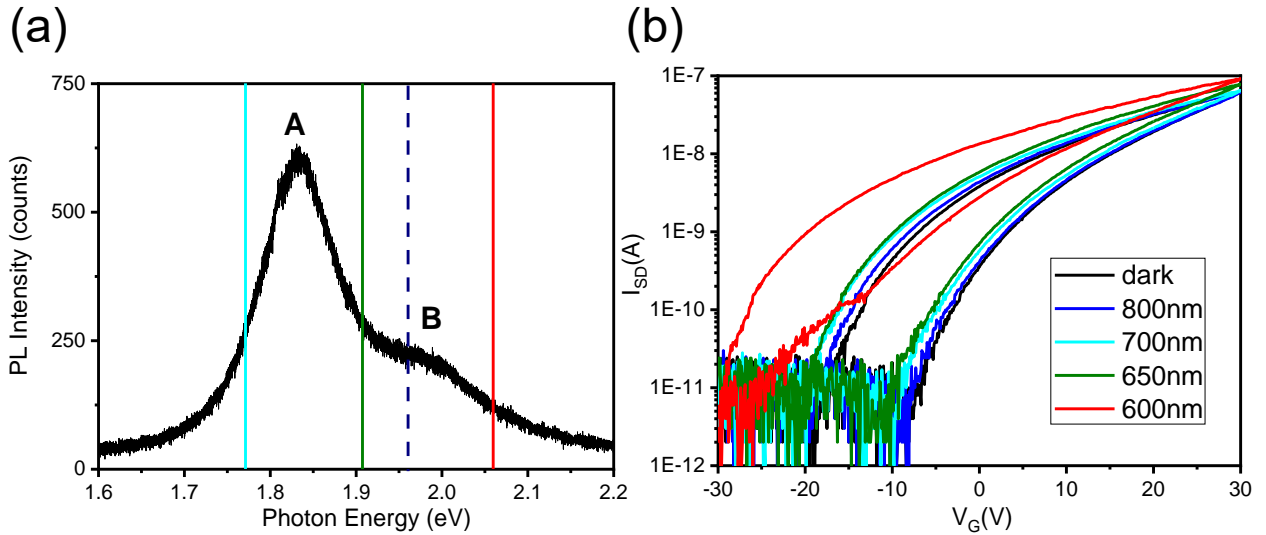

**Figure 5.** (a) Photoluminescence spectrum of $MoS_2$ on $Al_2O_3$. The vertical lines correspond to the photon energies for illumination with different wavelengths (700 nm, 650 nm and 600 nm for the solid lines and 633 nm for the dashed line). (b) The $I_{SD} - V_G$ curves of $MoS_2$ device on $Al_2O_3$ under illumination with different wavelengths.

In conclusion, we studied how the photoresponse of $MoS_2$ phototransitors is affected by ambient conditions. We find that illumination with photon energy sufficiently high to excite electron-hole pairs causes the desorption of water and oxygen molecules from the $MoS_2$ surface, due the increased number of photogenerated holes. This photogating effect dominates the photoresponse in ambient conditions and sets the response time, which is determined by the absorption and desorption processes of molecules on the $MoS_2$. The response time of $MoS_2$ photodetectors in ambient conditions (tens of seconds or higher) is therefore very similar to the response time of $MoS_2$ gas sensors [24]. Altough photogating due to the molecular gating of oxygen and water provides very high responsivity, it is not suitable for fast detection. Laser illumination and vacuum pumping combined with the use of hydrophobic substrates are effective methods to remove absorbates before device passivation, to obtain two dimensional devices with small hysteresis and fast photoresponse.

METHODS

*Tunable Light Source*: We combined a Xenon light source with a monochromator to tune the wavelength of the output light from 600 nm to 800 nm. The output of the lamp was collimated and then focused onto the sample using a 20X long-working-distance objective lens. The power output in the wavelength range from 600 nm to 800 nm varied between 1 μW and 2 μW, the diameter of the light spot focused on the device was about 60 μm and the width between the two electrodes was 12 μm. A CCD camera was used to monitor the position of the light spot.

AUTHOR INFORMATION

**Corresponding Authors**

* Paola.Barbara@georgetown.edu; * ph523@georgetown.edu.

**Present Addresses**

† Current Address: Ecole Centrale Casablanca, Bouskoura, Ville Verte, 27182, Casablanca, Morocco.

**Notes**

The authors declare no competing financial interests.

ACKNOWLEDGEMENTS

This work was supported by the by the US Office of Naval Research (N00014-16-1-2674), the NSF (ECCS-1610953, MRI CHE-1429079) and the Georgetown Environmental Initiative. SM would like to thank the Walter G. Mayer Endowed Scholarship Fund for support.

REFERENCES

1. Buscema, M.; Island, J. O.; Groenendijk, D. J.; Blanter, S. I.; Steele, G. A.; van der Zant, H. S. J.; Castellanos-Gomez, A., Photocurrent generation with two-dimensional van der Waals semiconductors. *Chemical Society Reviews* **2015,** *44* (11), 3691-3718.
2. Kufer, D.; Konstantatos, G., Highly Sensitive, Encapsulated $MoS_2$ Photodetector with Gate Controllable Gain and Speed. *Nano Letters* **2015,** *15* (11), 7307-7313.
3. Han, P.; St Marie, L.; Wang, Q. X.; Quirk, N.; El Fatimy, A.; Ishigami, M.; Barbara, P., Highly sensitive $MoS_2$ photodetectors with graphene contacts. *Nanotechnology* **2018,** *29* (20).
4. Lopez-Sanchez, O.; Lembke, D.; Kayci, M.; Radenovic, A.; Kis, A., Ultrasensitive photodetectors based on monolayer $MoS_2$. *Nature Nanotechnology* **2013,** *8* (7), 497-501.
5. Zhang, W. J.; Huang, J. K.; Chen, C. H.; Chang, Y. H.; Cheng, Y. J.; Li, L. J., High-Gain Phototransistors Based on a CVD $MoS_2$ Monolayer. *Advanced Materials* **2013,** *25* (25), 3456-3461.
6. Late, D. J.; Liu, B.; Matte, H.; Dravid, V. P.; Rao, C. N. R., Hysteresis in Single-Layer $MoS_2$ Field Effect Transistors. *Acs Nano* **2012,** *6* (6), 5635-5641.
7. Cho, K.; Park, W.; Park, J.; Jeong, H.; Jang, J.; Kim, T. Y.; Hong, W. K.; Hong, S.; Lee, T., Electric Stress-Induced Threshold Voltage Instability of Multilayer $MoS_2$ Field Effect Transistors. *Acs Nano* **2013,** *7* (9), 7751-7758.

8. Qiu, H.; Pan, L. J.; Yao, Z. N.; Li, J. J.; Shi, Y.; Wang, X. R., Electrical characterization of back-gated bi-layer MoS$_2$ field-effect transistors and the effect of ambient on their performances. *Applied Physics Letters* **2012,** *100* (12).
9. Park, W.; Park, J.; Jang, J.; Lee, H.; Jeong, H.; Cho, K.; Hong, S.; Lee, T., Oxygen environmental and passivation effects on molybdenum disulfide field effect transistors. *Nanotechnology* **2013,** *24* (9).
10. Ahn, J. H.; Parkin, W. M.; Naylor, C. H.; Johnson, A. T. C.; Drndic, M., Ambient effects on electrical characteristics of CVD-grown monolayer MoS$_2$ field-effect transistors. *Scientific Reports* **2017,** *7*.
11. Pak, J.; Min, M.; Cho, K.; Lien, D. H.; Ahn, G. H.; Jang, J.; Yoo, D.; Chung, S.; Javey, A.; Lee, T., Improved photoswitching response times of MoS$_2$ field-effect transistors by stacking p-type copper phthalocyanine layer. *Applied Physics Letters* **2016,** *109* (18).
12. Di Bartolomeo, A.; Genovese, L.; Giubileo, F.; Iemmo, L.; Luongo, G.; Foller, T.; Schleberger, M., Hysteresis in the transfer characteristics of MoS$_2$ transistors. *2d Materials* **2018,** *5* (1).
13. Pospischil, A.; Mueller, T., Optoelectronic Devices Based on Atomically Thin Transition Metal Dichalcogenides. *Applied Sciences-Basel* **2016,** *6* (3).
14. Tselev, A.; Hatton, K.; Fuhrer, M. S.; Paranjape, M.; Barbara, P., A photolithographic process for fabrication of devices with isolated single-walled carbon nanotubes. *Nanotechnology* **2004,** *15* (11), 1475-1478.
15. Nan, H. Y.; Wang, Z. L.; Wang, W. H.; Liang, Z.; Lu, Y.; Chen, Q.; He, D. W.; Tan, P. H.; Miao, F.; Wang, X. R.; Wang, J. L.; Ni, Z. H., Strong Photoluminescence Enhancement of MoS$_2$ through Defect Engineering and Oxygen Bonding. *Acs Nano* **2014,** *8* (6), 5738-5745.
16. Furchi, M. M.; Polyushkin, D. K.; Pospischil, A.; Mueller, T., Mechanisms of Photoconductivity in Atomically Thin MoS$_2$. *Nano Letters* **2014,** *14* (11), 6165-6170.
17. Di Bartolomeo, A.; Genovese, L.; Foller, T.; Giubileo, F.; Luongo, G.; Croin, L.; Liang, S. J.; Ang, L. K.; Schleberger, M., Electrical transport and persistent photoconductivity in monolayer MoS$_2$ phototransistors. *Nanotechnology* **2017,** *28* (21).
18. Tongay, S.; Zhou, J.; Ataca, C.; Liu, J.; Kang, J. S.; Matthews, T. S.; You, L.; Li, J. B.; Grossman, J. C.; Wu, J. Q., Broad-Range Modulation of Light Emission in Two-Dimensional Semiconductors by Molecular Physisorption Gating. *Nano Letters* **2013,** *13* (6), 2831-2836.
19. Miller, B.; Parzinger, E.; Vernickel, A.; Holleitner, A. W.; Wurstbauer, U., Photogating of mono- and few-layer MoS$_2$. *Applied Physics Letters* **2015,** *106* (12).
20. Radisavljevic, B.; Radenovic, A.; Brivio, J.; Giacometti, V.; Kis, A., Single-layer MoS$_2$ transistors. *Nature Nanotechnology* **2011,** *6* (3), 147-150.
21. Wang, G.; Chernikov, A.; Glazov, M. M.; Heinz, T. F.; Marie, X.; Amand, T.; Urbaszek, B., Colloquium: Excitons in atomically thin transition metal dichalcogenides. *Reviews of Modern Physics* **2018,** *90* (2).
22. Rigosi, A. F.; Hill, H. M.; Rim, K. T.; Flynn, G. W.; Heinz, T. F., Electronic band gaps and exciton binding energies in monolayer Mo$_x$W$_{1-x}$S$_2$ transition metal dichalcogenide alloys probed by scanning tunneling and optical spectroscopy. *Physical Review B* **2016,** *94* (7).
23. Fang, H. H.; Hu, W. D., Photogating in Low Dimensional Photodetectors. *Advanced Science* **2017,** *4* (12).
24. Kumar, R.; Goel, N.; Kumar, M., UV-Activated MoS2 Based Fast and Reversible NO$_2$ Sensor at Room Temperature. *Acs Sensors* **2017,** *2* (11), 1744-1752.